\documentclass[12pt]{article}
\usepackage{amsmath}
\usepackage{amssymb}
\usepackage{citesort}
\usepackage{graphicx}
\usepackage[nomarkers]{endfloat}
\usepackage{pstricks}
\numberwithin{equation}{section}
\begin{document}

\setcounter{page}{0}
\thispagestyle{empty}

\begin{flushright}
{\small BARI-TH 373/00}
\end{flushright}

\vspace*{2.5cm}

\begin{center}
{\large\bf A gauge invariant study of the monopole condensation in  \\[0.4cm]
non Abelian lattice gauge theories}
\end{center}

\vspace*{2cm}

\renewcommand{\thefootnote}{\fnsymbol{footnote}}

\begin{center}
{
Paolo Cea$^{1,2,}$\protect\footnote{Electronic address:
{\tt Paolo.Cea@bari.infn.it}} and
Leonardo Cosmai$^{2,}$\protect\footnote{Electronic address:
{\tt Leonardo.Cosmai@bari.infn.it}} \\[0.5cm]
$^1${\em Dipartimento di Fisica, Universit\`a di Bari,
I-70126 Bari, Italy}\\[0.3cm]
$^2${\em INFN - Sezione di Bari,
I-70126 Bari, Italy}
}
\end{center}

\vspace*{0.5cm}

\begin{center}
{
June, 2000
}
\end{center}

\vspace*{1.0cm}

\renewcommand{\abstractname}{\normalsize Abstract}
\begin{abstract}
We investigate the Abelian monopole condensation in finite
temperature SU(2) and SU(3) pure lattice gauge theories. To this
end we introduce a gauge invariant disorder parameter built up in
terms of the lattice Schr\"odinger functional. Our numerical
results show that the disorder parameter is different from zero
and Abelian monopole condense in the confined phase. On the other
hand our numerical data suggest that the disorder parameter tends
to zero, in the thermodynamic limit, when the gauge coupling
constant approaches the critical deconfinement value. In
the case of SU(3) we also compare the different kinds of Abelian
monopoles which can be defined according to the choice of the
Abelian subgroups.
\end{abstract}

\vspace*{0.5cm}
\begin{flushleft}
PACS number(s): 11.15.Ha
\end{flushleft}
\renewcommand{\thesection}{\normalsize{\Roman{section}.}}
\section{\normalsize{INTRODUCTION}}
\renewcommand{\thesection}{\arabic{section}}

The dual superconductivity of the vacuum in gauge theories to
explain color confinement has been proposed since long time by
G. 't Hooft~\cite{tHooft:1975} and
S. Mandelstam~\cite{Mandelstam:1976}.
These authors proposed that the confining vacuum behaves as a
coherent state of color magnetic monopoles. In other words the
confining vacuum is a magnetic (dual) superconductor. This
fascinating proposal offers a picture of confinement whose physics
can be clearly extracted. Indeed, the dual Meissner effect causes
the formation of chromoelectric flux tubes between chromoelectric
charges leading to a linear confining potential.

Following Ref.~\cite{tHooft:1981} let us consider gauge theories
without matter fields. In order to realize gauge field
configurations which describe magnetic monopoles we need a scalar
Higgs field~\cite{Polyakov:1974}. In the 't Hooft's scheme the
role of the scalar field is played by any operator which
transforms in the adjoint representation of the gauge group. Let
$X(x)$ be an operator in the adjoint representation, then one
fixes the gauge by diagonalizing $X(x)$ at each point. This choice
does not fix completely the gauge, for it leaves as residual
invariance group the maximal Abelian (Cartan) subgroup of the
gauge group. This procedure is known as Abelian
projection~\cite{tHooft:1981}. The world line of the monopoles can
be identified as the lines where two eigenvalues of the operator
$X(x)$ are equal. Thus, the dual superconductor idea is realized
if these Abelian monopole condense. Due to the gauge invariance we
expect that the monopole condensation should manifest irrespective
to the gauge fixing. In other words all the Abelian projections
are physically equivalent. However, it is conceivable that the
dual superconductor scenario could manifest clearly with a clever
choice of the operator $X(x)$. It is remarkable that, if one
adopts the so called maximally Abelian
projection~\cite{Kronfeld:1987}, then it seems that the Abelian
projected links retain the information relevant to the
confinement~\cite{Suzuki:1993}.

It turns out that the Abelian projection can be implemented on the
lattice~\cite{Kronfeld:1987}, so that one can analyze the dynamics
of the Abelian projected gauge fields by means of non perturbative
numerical simulations.
Indeed, the first direct evidence of the dual Abrikosov vortex
joining two static quark-antiquark  pair has been obtained in
lattice simulations of gauge
theories~\cite{Suzuki:1993,Singh:1993,Cea:1993,Bali:1995}. In
particular in Ref.~\cite{Cea:1993} we considered the pure gauge
$SU(2)$ lattice theory and found evidence of the dual Meissner
effect both in the maximally Abelian gauge and without gauge
fixing. Moreover we showed that the London penetration length is a
physical gauge invariant quantity.

An alternative and more direct method to detect the dual
superconductivity relies upon the very general assumption that the
dual superconductivity of the ground state is realized if there is
condensation of Abelian monopoles. Thus, according to
Ref.~\cite{DiGiacomo:1994} it suffices to measure a disorder
parameter defined as the vacuum expectation value of a nonlocal
operator with non zero magnetic charge and non vanishing vacuum
expectation value in the confined phase. However, in the case of
non Abelian gauge theories, the disorder parameter is expected to
break a non Abelian symmetry, while the dual superconductivity is
realized by condensation of Abelian monopoles. As we have already
argued, the Abelian monopole charge can be associated to each
operator in the adjoint representation by the so-called Abelian
projection~\cite{tHooft:1981,Kronfeld:1987}. Indeed, the authors
of Ref.~\cite{DiGiacomo:1994} introduced on the lattice a disorder
parameter describing condensation of monopoles within a particular
Abelian projection. On the other hand, recent
results~\cite{DiGiacomo:2000} show that the Abelian monopoles
defined through several Abelian projection condense, suggesting
that the monopole condensation does not depend on the adjoint
operator used in the Abelian projection procedure. This is in
accordance with the theoretical expectation that monopole
condensation should occur irrespective of the gauge fixing
procedure. However, a gauge invariant evidence of the Abelian
monopole condensation is still lacking.

The aim of the present paper is to investigate the Abelian
monopole condensation in pure lattice gauge SU(2) and SU(3)
theories in a gauge-invariant way~\cite{Cea:1999}. To do this we
introduce a disorder parameter defined in terms of a
gauge-invariant thermal partition functional in presence of an
external background field. \\
The plan of the paper is as follows. In
Section~II we introduce the thermal partition functional, built up
using the lattice Schr\"odinger
functional~\cite{Rossi:1980,Luescher:1992}. In Section~III we
study the Abelian monopole condensation for finite temperature
SU(2) lattice gauge theory. Section~IV is devoted to the case of
SU(3) gauge theory at finite temperature, where, according to the
choice of the Abelian subgroup, different kinds of Abelian
monopoles can be defined. Our conclusions are drawn in Section~V.

\renewcommand{\thesection}{\normalsize{\Roman{section}.}}
\section{\normalsize{THE THERMAL PARTITION FUNCTIONAL}}
\renewcommand{\thesection}{\arabic{section}}

To investigate the dynamics of the vacuum at zero temperature
we introduced~\cite{metodo,PRD} the gauge-invariant effective action
for external static (i.e. time-independent) background field
defined by means of the lattice Schr\"odinger functional:
\begin{equation}
\label{Zetalatt}
{\mathcal{Z}}[U^{\mathrm{ext}}_\mu] = \int {\mathcal{D}}U \; e^{-S_W} \,,
\end{equation}
where $S_W$ is the standard Wilson action. The functional
integration is extended over links on a lattice with the
hypertorus geometry and satisfying the constraints
\begin{equation}
\label{coldwall}
U_\mu(x)|_{x_4=0} = U^{\mathrm{ext}}_\mu(\vec{x})  \,.
\end{equation}
In Equations~(\ref{Zetalatt}) and (\ref{coldwall})
$U^{\mathrm{ext}}_\mu(\vec{x})$ is the lattice version of
the external continuum gauge field
$\vec{A}^{\mathrm{ext}}(\vec{x})=\vec{A}^{\mathrm{ext}}_a(\vec{x}) \lambda_a/2$:
\begin{equation}
\label{links}
U_\mu^{\text{ext}}(\vec{x}) = {\mathrm P} \exp\left\{
iag \int_0^1 dt \, A^{\text{ext}}_{a,\mu}(\vec{x}+t\hat{\mu}) \frac{\lambda_a}{2} \right\} \,,
\end{equation}
where ${\mathrm P}$ is the path-ordering operator
and $g$ the gauge coupling constant. \\
The lattice effective action for the external static background field
$\vec{A}^{\mathrm{ext}}(\vec{x})$ is given by
\begin{equation}
\label{Gamma}
\Gamma[\vec{A}^{\mathrm{ext}}] = -\frac{1}{L_4} \ln \left\{
\frac{{\mathcal{Z}}[\vec{A}^{\mathrm{ext}}]}{{\mathcal{Z}}(0)} \right\} \,,
\end{equation}
where $L_4$ is the extension in Euclidean time and
$\mathcal{Z}(0)$ is the lattice Schr\"odinger functional,
Eq.~(\ref{Zetalatt}), without the external background field
($U^{\mathrm{ext}}_\mu = \mathbf{1}$).
It can be shown~\cite{metodo} that in the continuum limit
$\Gamma[\vec{A}^{\mathrm{ext}}]$ is the vacuum energy in presence
of the background field $\vec{A}^{\mathrm{ext}}(\vec{x})$.

We want now to extend our definition of lattice
effective action to gauge systems at finite temperature.
In this case the relevant
quantity is the thermal partition function.
In the continuum we have:
\begin{equation}
\label{thermal}
\text{Tr}\left[ e^{-\beta_T H} \right] =
\int \mathcal{D}\vec{A} \,
\langle \vec{A} \left | e^{-\beta_T H} \mathcal{P} \right|
\vec{A} \rangle \,,
\end{equation}
where $\beta_T$ is the inverse of the physical temperature,
$H$ is the Hamiltonian, and $\mathcal{P}$ projects onto the
physical states. As is well known, the thermal partition function
can be written as~\cite{Gross:1981}:
\begin{equation}
\label{tpf}
\text{Tr}\left[ e^{-\beta_T H} \right] =
\int_{A_\mu(\beta_T,\vec{x})=A_\mu(0,\vec{x})}
 \mathcal{D}A_\mu(x_4,\vec{x})  \,
e^{-\int^{\beta_T}_0 dx_4 \, \int d^3 \vec{x} \mathcal{L}_{Y-M}(\vec{x},x_4)}
\,.
\end{equation}
On the lattice we have:
\begin{equation}
\label{tpflattice}
\text{Tr}\left[ e^{-\beta_T H} \right] =
\int_{U_\mu(\beta_T,\vec{x})=U_\mu(0,\vec{x})=U_\mu(\vec{x})}
 \mathcal{D}U_\mu(x_4,\vec{x})  \,
e^{-S_W} \,.
\end{equation}
Comparing Eq.~(\ref{tpflattice}) with Eqs.~(\ref{Zetalatt})
and~(\ref{coldwall}), we get:
\begin{equation}
\label{trace}
\text{Tr}\left[ e^{-\beta_T H} \right] =
\int \mathcal{D}U_\mu(\vec{x})  \,
\mathcal{Z}[U_\mu(\vec{x})] \,,
\end{equation}
where $\mathcal{Z}[U_\mu(\vec{x})]$ is the Schr\"odinger functional
Eq.~(\ref{Zetalatt}) defined on a lattice with $L_4=\beta_T$,
with ``external'' links $U_\mu(\vec{x})$ at $x_4=0$.

We are interested in the thermal partition function in presence
of a given static background field $\vec{A}^{\mathrm{ext}}(\vec{x})$.
In the continuum this can be obtained by splitting the
gauge field into the background field $\vec{A}^{\mathrm{ext}}(\vec{x})$
and the fluctuating fields $\eta(x)$. So that we could write formally
for the thermal partition function $\mathcal{Z}_T[\vec{A}^{\mathrm{ext}}]$:
\begin{equation}
\label{ZetaT}
\mathcal{Z}_T[\vec{A}^{\mathrm{ext}}] =
\int \mathcal{D} \vec{\eta} \,
\langle \vec{A}^{\mathrm{ext}}, \vec{\eta} \left|
e^{-\beta_T H} \mathcal{P} \right|
\vec{A}^{\mathrm{ext}}, \vec{\eta} \rangle \,.
\end{equation}
The lattice implementation of Eq.~(\ref{ZetaT})
can be obtained from Eq.~(\ref{tpflattice}) if we write
\begin{equation}
\label{ukbetat}
U_k(\beta_T,\vec{x})=U_k(0,\vec{x})=U^{\text{ext}}_k(\vec{x})
\widetilde{U}_k(\vec{x}) \,,
\end{equation}
where $U^{\text{ext}}_k(\vec{x})$ is given by Eq.~(\ref{links}) and the
$\widetilde{U}_k(\vec{x})$'s are the fluctuating links.
Thus we get
\begin{equation}
\label{zzz} \mathcal{Z}_T[\vec{A}^{\mathrm{ext}}] =
\int \,  \mathcal{D}\widetilde{U}_k(\vec{x}) \,
\mathcal{D}{U}_4(\vec{x}) \,
\mathcal{Z}[U_k^{\text{ext}}(\vec{x}),\widetilde{U}_k(\vec{x})]
\,,
\end{equation}
where we integrate over the fluctuating links $\widetilde{U}_k(\vec{x})$, while
the $U_k^{\text{ext}}$ links are fixed.
Note that
in Eq.~(\ref{zzz}) only the spatial links belonging to the
hyperplane $x_4=0$ are written as the product of the external link
$U^{\text{ext}}_k(\vec{x})$ and the fluctuating links
$\widetilde{U}_k(\vec{x})$. The temporal links $U_4(x_4=0,\vec{x})$
are left freely fluctuating. It follows that the temporal links
$U_4(x)$ satisfy the usual periodic boundary conditions.
We stress that the periodic boundary conditions in the temporal
direction are crucial to retain the physical interpretation that
the functional $\mathcal{Z}_T[\vec{A}^{\text{ext}}]$ is a
thermal partition function. In the following the
spatial links belonging to the time-slice $x_4=0$ will
be called ``frozen links'', while the remainder will be
the ``dynamical links''. \\
From the physical point of view
we are considering the gauge system at finite temperature
in interaction with a fixed external background field. As
a consequence, in the Wilson action $S_W$ we keep only the plaquettes
built up with the dynamical links or with dynamical and frozen
links. With these limitations it is easy to see that
in Eq.~(\ref{zzz}) we have
\begin{equation}
\label{ZetaText}
\mathcal{Z} \left[ U_k^{\text{ext}}(\vec{x}),
\widetilde{U}_k(\vec{x}) \right]
= \mathcal{Z} \left[U_k^{\text{ext}}(\vec{x}) \right] \,.
\end{equation}
Indeed, let us consider an arbitrary frozen link
$U^{\text{ext}}_k(\vec{x}) \widetilde{U}_k(\vec{x})$.
This link enters in the modified Wilson action by means of
the plaquette:
\begin{equation}
\label{plaquette}
P_{k4}(x_4=0,\vec{x}) = \text{Tr} \left\{
U^{\text{ext}}_k(\vec{x})  \widetilde{U}_k(\vec{x})
U_4(0,\vec{x}+\hat{k}) U^\dagger_k(1,\vec{x})
U^\dagger_4(0,\vec{x}) \right\} \,.
\end{equation}
Now we observe that the link $U_4(0,\vec{x}+\hat{k})$
in Eq.~(\ref{plaquette}) is a dynamical one, i.e. we are
integrating over it. So that, by using the invariance
of the Haar measure we obtain
\begin{equation}
\label{plaqnew}
P_{k4}(x_4=0,\vec{x}) = \text{Tr} \left\{
U^{\text{ext}}_k(\vec{x})
U_4(0,\vec{x}+\hat{k}) U^\dagger_k(1,\vec{x})
U^\dagger_4(0,\vec{x}) \right\} \,.
\end{equation}
It is evident that Eq.~(\ref{plaqnew}) in turns implies
Eq.~(\ref{ZetaText}). Then, we see that in Eq.~(\ref{zzz})
the integration over the fluctuating links $\widetilde{U}(\vec{x})$
gives an irrelevant multiplicative constant.
So that we have:
\begin{equation}
\label{ZetaTnew}
\mathcal{Z}_T \left[ \vec{A}^{\text{ext}} \right] =
\int_{U_k(\beta_T,\vec{x})=U_k(0,\vec{x})=U^{\text{ext}}_k(\vec{x})}
\mathcal{D}U \, e^{-S_W}   \,,
\end{equation}
where the integrations are over the dynamical links with
periodic boundary conditions in the time direction.
As concerns the boundary conditions at the spatial boundaries,
we keep the fixed boundary conditions 
$U_k(\vec{x},x_4)=U_k^{\text{ext}}(\vec{x})$
used in the Schr\"odinger functional Eq.(\ref{Zetalatt}).
Thus we see that, if we send the
physical temperature to zero, then the thermal functional
Eq.~(\ref{ZetaTnew}) reduces to the zero-temperature
Schr\"odinger functional Eq.~(\ref{Zetalatt}) with the constraints
$U_k(x)|_{x_4=0} = U^{\mathrm{ext}}_k(\vec{x})$ instead of
Eq.~(\ref{coldwall}).
In our previous study~\cite{metodo}
we checked that in the thermodynamic limit
both conditions agree as concerns
the zero-temperature effective action Eq.~(\ref{Gamma}).

\renewcommand{\thesection}{\normalsize{\Roman{section}.}}
\section{\normalsize{ABELIAN MONOPOLE CONDENSATION: SU(2)}}
\renewcommand{\thesection}{\arabic{section}}

Let us consider the SU(2) pure gauge theory at finite temperature.
We are interested in the thermal partition function
Eq.~(\ref{ZetaTnew}) in presence of an Abelian monopole field.
In the case of SU(2) gauge theory the maximal Abelian group is
an Abelian U(1) group. Thus, in the continuum the Abelian monopole
field turns out to be:
\begin{equation}
\label{monop3}
g \vec{b}^a({\vec{x}}) = \delta^{a,3} \frac{n_{\mathrm{mon}}}{2}
\frac{ \vec{x} \times \vec{n}}{|\vec{x}|(|\vec{x}| - \vec{x}\cdot\vec{n})} \,.
\end{equation}
where $\vec{n}$ is the direction of the Dirac string and, according
to the Dirac quantization condition, $n_{\text{mon}}$ is an integer.
The lattice links corresponding to the Abelian monopole field
Eq.~(\ref{monop3}) can be readily obtained as:
\begin{equation}
\label{latlinks}
U^{\text{ext}}_k(\vec{x}) =
\text{P} \exp \left\{
ig \, \int_0^1 dt \,  \frac{\sigma_a}{2}b^a_k(\vec{x} + t \hat{x}_k)
\right\} \,,
\end{equation}
where the $\sigma_a$'s are the Pauli matrices.
By choosing $\vec{n}=x_3$ we get:
\begin{equation}
\label{su2links}
\begin{split}
U^{\text{ext}}_{1,2}(\vec{x})  & =
\cos [ \theta_{1,2}(\vec{x}) ] +
i \sigma_3 \sin [ \theta_{1,2}(\vec{x}) ] \,, \\
U^{\text{ext}}_{3}(\vec{x}) & =  {\mathbf 1} \,,
\end{split}
\end{equation}
with
\begin{equation}
\label{thetat3}
\begin{split}
\theta_1(\vec{x}) & = -\frac{n_{\text{mon}}}{4}
\frac{(x_2-X_2)}{|\vec{x}_{\text{mon}}|}
\frac{1}{|\vec{x}_{\text{mon}}| - (x_3-X_3)} \,, \\
\theta_2(\vec{x}) & = +\frac{n_{\text{mon}}}{4}
\frac{(x_1-X_1)}{|\vec{x}_{\text{mon}}|}
\frac{1}{|\vec{x}_{\text{mon}}| - (x_3-X_3)} \,.
\end{split}
\end{equation}
In Equation~(\ref{thetat3}) $(X_1,X_2,X_3)$ are the monopole coordinates
and $\vec{x}_{\text{mon}} = (\vec{x} - \vec{X})$.
In the numerical simulations we put the lattice Dirac monopole
at the center of the time slice $x_4=0$. To avoid
the singularity due to the Dirac string we locate the monopole
between two neighboring sites. We have checked
that the numerical results are not too sensitive to
the precise position of the magnetic monopole.

According to the discussion in the previous Section we are interested in
the thermal partition function $\mathcal{Z}_T[\vec{A}^{\text{ext}}]$
given by Eq.~(\ref{ZetaTnew}). Note that we do not need to fix the
gauge due to the gauge invariance of the thermal partition
functional against gauge transformations of the external
background field.
On the lattice the physical temperature $T_{\text{phys}}$
is given by
\begin{equation}
\label{Tphys}
\frac{1}{T_{\text{phys}}} = \beta_T = L_t \,,
\end{equation}
where $L_t = L_4$ is the lattice linear extension in the
time direction. In order to approximate the thermodynamic limit
the spatial extension $L_s$ should satisfy
\begin{equation}
\label{lsgglt}
L_s \gg L_t \,.
\end{equation}
To this end we performed our numerical simulations
on lattices such that
\begin{equation}
\label{ltls4}
\frac{L_t}{L_s} \le 4 \,.
\end{equation}
In the numerical simulations we impose periodic boundary conditions in the
time direction. As already discussed, at the spatial boundaries the links
are fixed according to Eq.~(\ref{su2links}). This last condition
corresponds to the requirement that the fluctuations over
the background field vanish at infinity.

Following the suggestion of Ref.~\cite{DiGiacomo:2000} we introduce
the gauge-invariant disorder parameter for confinement
\begin{equation}
\label{disorder}
\mu = e^{-F_{\text{mon}}/T_{\text{phys}}} =
\frac{\mathcal{Z}_T[n_{\text{mon}}]}{\mathcal{Z}_T[0]} \,,
\end{equation}
where $\mathcal{Z}_T[0]$ is the thermal partition function
without monopole field (i.e. with $n_{\text{mon}} = 0$).

From Eq.~(\ref{disorder}) it is clear that $F_{\text{mon}}$
is the free energy to create an Abelian monopole.
If there is monopole condensation, then $F_{\text{mon}}=0$
and $\mu = 1$.
To avoid the problem of measuring a partition function we focus on
the derivative of the monopole free energy:
\begin{equation}
\label{derivative}
F^\prime_{\text{mon}} =
\frac{\partial}{\partial \beta} F_{\text{mon}} \,.
\end{equation}
It is straightforward to see that $F^\prime_{\text{mon}}$
is given by the difference between the average plaquette
\begin{equation}
\label{avplaq}
F^\prime_{\text{mon}} = V \left[ <P>_{n_{\text{mon}}=0} -
<P>_{n_{\text{mon}} \ne 0} \right] \,,
\end{equation}
where $V$ is the spatial volume.

We use the over-relaxed heat-bath algorithm to update the gauge configurations.
Simulations have been performed by means of the APE100/Quadrics computer facility in Bari.
Since we are measuring a local quantity such as the plaquette, a low statistics
(from 2000 up to 12000 configurations) is required in order to get a good estimation
of $F^{\prime}_{\mathrm{mon}}$.

In Figure~1 we display the derivative of the monopole free energy
versus $\beta$ for $n_{\text{mon}}=10$ on a lattice with $L_t =4$
and $L_s = 24$. We see that $F^\prime_{\text{mon}}$ vanishes at
strong coupling and displays a rather sharp peak near $\beta
\backsimeq 2.13 $. We expect that the peak corresponds to the
finite temperature deconfinement transition. In   Figure~1 we also
display the absolute value of the Polyakov loop:
\begin{equation}
\label{abspolysu2}
| P | = < |  
\frac{1}{V} \sum_{\vec{x}} \frac{1}{2} \mathrm{Tr} [ 
\prod_{t=1}^{L_t} U_4(\vec{x},t) ] | > \,,
\end{equation}
and, indeed, we see that the peak corresponds to the rise of
Polyakov loop.\\
In the weak coupling region the plateau in
$F^\prime_{\text{mon}}$ indicates that the monopole free energy
tends to the classical monopole action which behaves linearly in
$\beta$. To see this, we observe that deeply in the weak coupling
region the lattice action should reduce to the classical action.
In the naive continuum limit the classical action reads :
\begin{equation}
\label{classicalsu2}
 S_{\text{class}} = \; \frac{1}{2} \; \int^{\beta_T}_0 dx_4 \,
 \int d^3 \vec{x} \vec{B}^a(\vec{x}) \; \vec{B}^a(\vec{x})    \;
\end{equation}
where $\vec{B}^a(\vec{x})$ is the classical Abelian monopole
magnetic field. Introducing an ultraviolet cutoff
 $\Lambda = \alpha/a$, with $\alpha$ a constant and $a$ the
lattice spacing, and performing in  Eq.~(\ref{classicalsu2}) the
spatial integral over the volume $V = L_s^3$, we get:
\begin{equation}
\label{classical2}
 S_{\text{class}} \backsimeq \; \frac{\pi \alpha \beta}{8 \, T_{\mathrm{phys}}} \;
 n^{2}_{\text{mon}}  + \text{O}(1/L_s a) \; .
\end{equation}
So that in the weak coupling region we have :
\begin{equation}
\label{fprimeclas}
F^{\prime}_{\mathrm{mon}} \backsimeq \frac{\pi}{8} \alpha n^{2}_{\text{mon}}
 \; .
\end{equation}
From Figure~1 we see that Eq.~(\ref{fprimeclas}) with $\alpha
\backsimeq 1.2 $ ( dashed line) describes quite well the numerical
data in the relevant region. \\
In order to determine the critical parameters and the order of the
transition, we need to perform the finite size scaling analysis.
We plan to do this in a future work. In this paper we restrict
ourself to a preliminary qualitative analysis. In Figure~2 we
compare the derivative of the monopole free energy  on lattices
with $L_t =4$ and $L_s = 24, 48$. We see that in the strong
coupling  $F^{\prime}_{\mathrm{mon}}$
agrees for the two lattices. On the other hand, in the weak
coupling region the different values of the plateaus 
can be ascribed to finite volume effects. In the critical
region we see that the peak increases. 
%

%
With the aim of obtaining the disorder parameter $\mu$
(Eq.~(\ref{disorder})) we perform the numerical integration of the
monopole free energy derivative
\begin{equation}
\label{numintegr}
F_{\text{mon}}(\beta) = \int_0^\beta d \beta^\prime \,
F^\prime_{\text{mon}}(\beta^\prime) \,.
\end{equation}
In Figure~3 we show the disorder parameter $\mu$ versus $\beta$
for lattices with $L_t =4$ and $L_s = 24, 48$. We see clearly
that $\mu=1$ in the confined phase. In other words the monopoles
condense in the vacuum. On the other hand, it seems that $\mu \to
0$ in the thermodynamic limit when $\beta$ reaches the critical
value . Indeed, by increasing the spatial volume of the lattice,
the disorder parameter $\mu$ decreases faster toward zero.
Moreover we see that the finite volume behavior of our disorder
parameter is consistent with a second order deconfinement phase
transition. \\
It worthwhile to comment on the finite volume effects. As a matter
of fact, it appears that, even though  the spatial volume of our
larger lattice looks enormous, we gain a rather small increase in
the peak value of the monopole free energy derivative. This can be
understood by observing that, due to our peculiar conditions at
the spatial boundaries, the dynamical volume is smaller than the
geometrical one. Moreover, it is well known that the fixed
boundary conditions  for the gauge fields lead to more severe
finite volume effects with respect to the usual periodic boundary
conditions. So that, to reach the thermodynamic limit we must
simulate our gauge system on lattices with very large spatial
volumes.
We stress again that the precise determination of the critical
parameters requires a finite size scaling which will be presented
elsewhere.

\renewcommand{\thesection}{\normalsize{\Roman{section}.}}
\section{\normalsize{ABELIAN MONOPOLE CONDENSATION: SU(3)}}
\renewcommand{\thesection}{\arabic{section}}

In the case of SU(3) gauge theory, the maximal Abelian group
is U(1)$\times$U(1). Therefore we have two different types
of Abelian monopole. Let us consider, firstly, the Abelian
monopole field given by Eq.~({\ref{monop3}}), which we call
the $T_3$ Abelian monopole.
The lattice links are given by
\begin{equation}
\label{t3links}
U_{1,2}^{\text{ext}}(\vec{x}) =
\begin{bmatrix}
e^{i \theta_{1,2}(\vec{x})} & 0 & 0 \\
0 &  e^{- i \theta_{1,2}(\vec{x})} & 0 \\
0 & 0 & 1
\end{bmatrix}
\,,
\end{equation}
with $\theta_{1,2}(\vec{x})$ defined in Eq.~(\ref{thetat3}).
The second type of independent Abelian monopole can be obtained by
considering the diagonal generator $T_8$.
In this case we have the  $T_8$ Abelian monopole:
\begin{equation}
\label{t8links} U_{1,2}^{\text{ext}}(\vec{x}) =
\begin{bmatrix}
e^{i \theta_{1,2}(\vec{x})} & 0 & 0 \\ 0 &  e^{i
\theta_{1,2}(\vec{x})} & 0 \\ 0 & 0 & e^{- 2 i
\theta_{1,2}(\vec{x})}
\end{bmatrix}
\,,
\end{equation}
with
\begin{equation}
\label{thetat8}
\begin{split}
\theta_1(\vec{x}) & = \frac{1}{\sqrt{3}} \left[
 -\frac{n_{\text{mon}}}{4}
\frac{(x_2-X_2)}{|\vec{x}_{\text{mon}}|}
\frac{1}{|\vec{x}_{\text{mon}}| - (x_3-X_3)} \right] \,, \\
\theta_2(\vec{x}) & =  \frac{1}{\sqrt{3}} \left[
+\frac{n_{\text{mon}}}{4} \frac{(x_1-X_1)}{|\vec{x}_{\text{mon}}|}
\frac{1}{|\vec{x}_{\text{mon}}| - (x_3-X_3)} \right] \,.
\end{split}
\end{equation}
Obviously, the lattice links Eq.~(\ref{t8links}) corresponds now
to the continuum gauge field
\begin{equation}
\label{T8monopole} g \vec{b}^a({\vec{x}}) = \delta^{a,8}
\frac{n_{\mathrm{mon}}}{2} \frac{ \vec{x} \times \vec{n}}{|\vec{x}|(|\vec{x}| -
\vec{x}\cdot\vec{n})} \,.
\end{equation}
The other Abelian monopoles can be generated by considering the
linear combination of the  $T_3$ and $T_8$ generators. In
particular we have considered the $T_{3a}$ Abelian monopole
corresponding to the following linear
combination~\cite{DiGiacomo:2000} of $\lambda_3/2$ and
$\lambda_8/2$:
\begin{equation}
\label{T3a} T_{3a} = -\frac{1}{2} \frac{\lambda_3}{2} +
\frac{\sqrt{3}}{2} \frac{\lambda_8}{2} =
\begin{bmatrix}
0 & 0 & 0 \\ 0 &  \frac{1}{2} & 0 \\ 0 & 0 & -\frac{1}{2}
\end{bmatrix}
\,.
\end{equation}
In Figure~4 we compare the free energy monopole derivative for the
$T_3$, $T_{3a}$ and $T_8$ Abelian monopoles for the lattice with
$L_t =4$ and $L_s = 32$.
We see that the $T_3$ and  $T_{3a}$ Abelian monopoles agree within
statistical errors in the whole range of $\beta$. On the other
hand the $T_8$ Abelian monopole displays a signal about a factor
two higher in the peak region. This is at variance of previous
studies~\cite{DiGiacomo:2000} which find out that the disorder
parameters for the three  Abelian monopoles defined by means of
the Polyakov projection coincide within statistical errors.
This result is quite interesting, for it suggests that in the
pattern of dynamical symmetry breaking due to the Abelian monopole
condensation the color direction $8$ is slightly preferred. \\
Let us consider, now, in detail the $T_8$ Abelian monopole. In
Figure~5 we report the derivative of the monopole free energy
versus $\beta$ for the lattice with $L_t =4$ and $L_s = 32$. We
also display the absolute value of the Polyakov loop:
\begin{equation}
\label{abspolysu3}
| P | = < |  
\frac{1}{V} \sum_{\vec{x}} \frac{1}{3} \mathrm{Tr} [ 
\prod_{t=1}^{L_t} U_4(\vec{x},t) ] | > \,,
\end{equation}
We see that $F^\prime_{\text{mon}}$ behaves like in the SU(2) case.
Indeed, the free energy monopole derivative is zero within errors
in the strong coupling region, while it display a sharp peak in
correspondence of the rise of the Polyakov loop. In the weak
coupling region $F^\prime_{\text{mon}}$ is almost constant. The
value of the plateau correspond to the classical action
Eq.~(\ref{classicalsu2}) which in the present case gives:
\begin{equation}
\label{classicalsu3}
 S_{\text{class}} \backsimeq \; \frac{\alpha \pi \beta}{12 \, T_{\mathrm{phys}}} \;
 n^{2}_{\text{mon}}  + \text{O}(1/L_s \text{a}) \; ,
\end{equation}
so that
\begin{equation}
\label{classical3}
F^\prime_{\mathrm{mon}} = \alpha \frac{\pi}{12} n^2_{\mathrm{mon}} \,.
\end{equation}
The dashed line in Figure~5 in the weak coupling region corresponds
to Eq.~(\ref{classical3}) with $\alpha \backsimeq 2.0$. \\
As in the $SU(2)$ theory we find that by increasing the spatial
volume the peak increases (see Figure~6). 
Our data do not show a measurable shift of the peak. 
We feel that this is a manifestation of the first order nature of the
$SU(3)$ deconfinement transition. This is confirmed if we look at
the disorder parameter $\mu$. In  Figure~7 we show the disorder
parameter $\mu$ versus $\beta$ for the  $L_t =4$ and $L_s = 32,
48$ lattices.
Again we see that  the disorder parameter $\mu$ is different from
zero in the confined phase and decreases towards zero in the
thermodynamic limit when we approach the critical coupling.
Moreover our numerical results suggest that by increasing the
spatial volume the two curves cross. This is precisely the finite
volume behavior expected for the order parameter in the case of a
first order phase transition~\cite{Ukawa:1990}.

\renewcommand{\thesection}{\normalsize{\Roman{section}.}}
\section{\normalsize{CONCLUSIONS}}
\renewcommand{\thesection}{\arabic{section}}

In this paper we have investigated the Abelian monopole
condensation in the finite temperature SU(2) and SU(3) lattice
gauge theories. By means of the lattice thermal partition
functional we introduce a disorder parameter which signals the
Abelian monopole condensation in the confined phase. By
construction our definition of the disorder parameter is gauge
invariant, so that we do not need to perform the Abelian
projection. Our numerical results suggest that the disorder
parameter $\mu$ is different from zero in the confined phase and
tends to zero when approaching the critical coupling in the
thermodynamic limit. We point out that in our approach the precise
determination of the critical parameters could be obtained by
means of a finite size scaling analysis. However, our results are
consistent with a second order  deconfining phase transition in
the case of the $SU(2)$ gauge theory. On the other hand, in the
case of $SU(3)$ the disorder parameter $\mu$ displays the
finite-size behavior expected for a first order transition. It is
clear that the finite size analysis in the critical region
requires a separate study with both better statistic and larger
lattice volumes. Remarkably, in the case of SU(3) gauge theory,
where there are two independent Abelian monopole fields related to
the two diagonal generators of the gauge algebra, we find that the
non perturbative vacuum reacts moderately strongly in the case of
the $T_8$ Abelian monopole. We feel that this last result should
be useful in the theoretical efforts to understand the pattern of
symmetry breaking in the deconfined phase of QCD. \\
In conclusion we stress that our approach, while keeping the gauge
invariance, can be readily extended to incorporate the dynamical
fermions. We hope to present results in this direction in a future
study.
%
%
%

\begin{figure}[H]
\label{Fig1}
\begin{center}
\includegraphics[clip,width=0.85\textwidth]{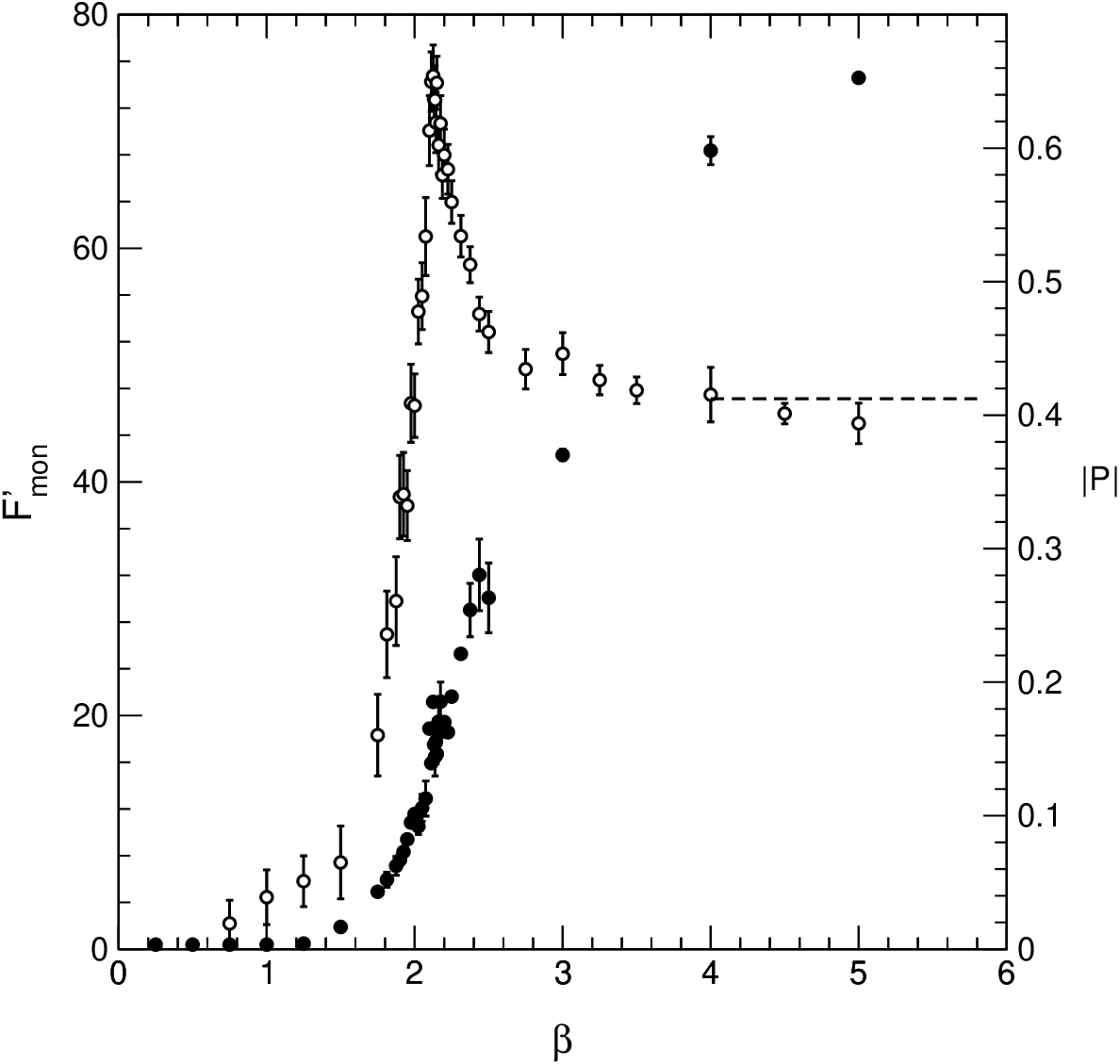}
\caption{The derivative of the SU(2) monopole free energy versus $\beta$ 
(Eq.~(\ref{avplaq}))
for $n_{\mathrm{mon}}=10$ on a lattice with $L_t=4$ and $L_s=24$ (open circles),
with the absolute value of the Polyakov loop (full circles). The dashed line
is Eq.~(\ref{classical2}).}
\end{center}
\end{figure}
\begin{figure}[H]
\label{Fig2}
\begin{center}
\includegraphics[clip,width=0.85\textwidth]{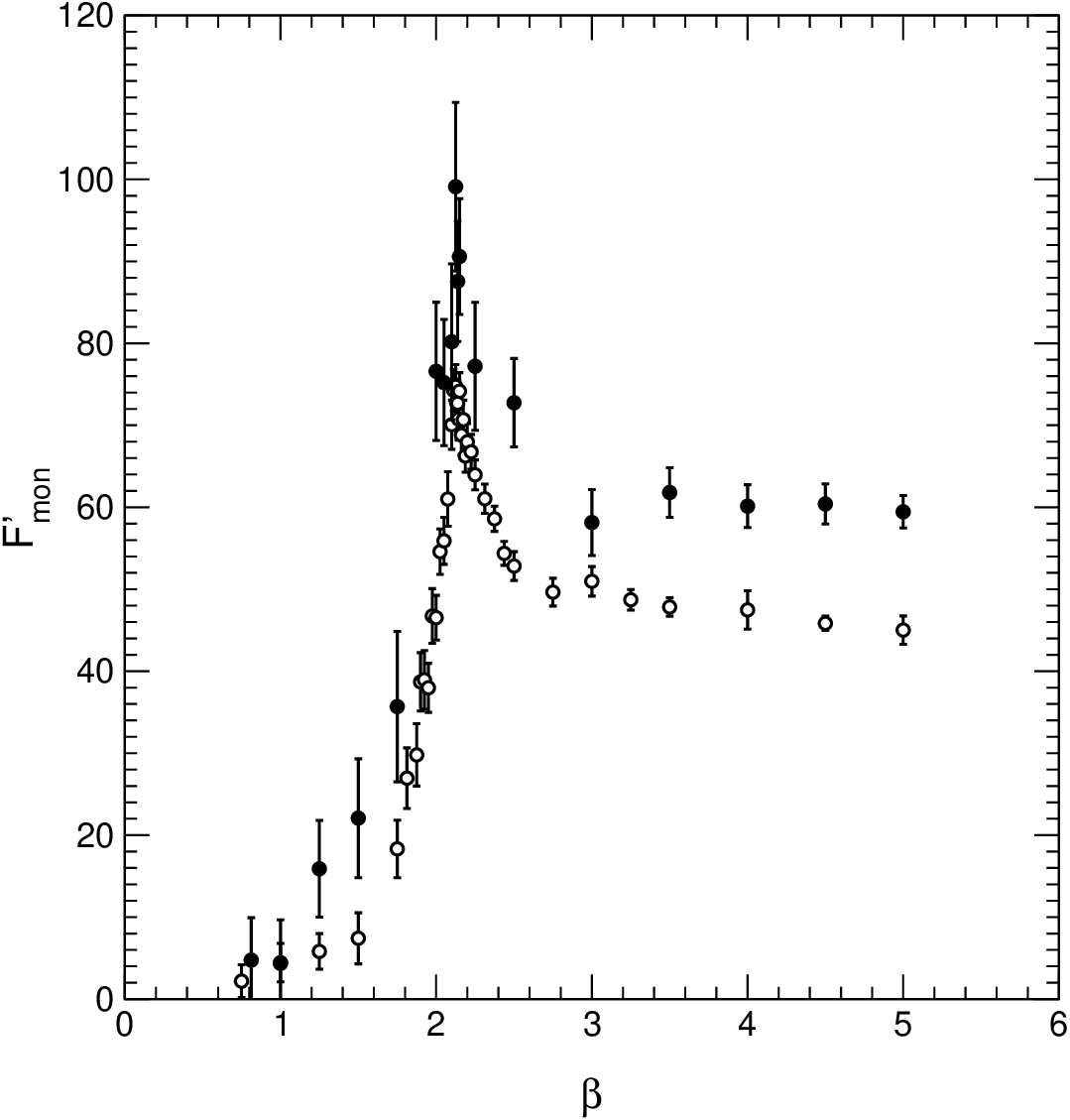}
\caption{The derivative of the SU(2) monopole free energy versus $\beta$,
(Eq.~(\ref{avplaq})), on lattices with $L_t=4$ and $L_s=24$ (open circles),
and $L_s=48$ (full circles).}
\end{center}
\end{figure}
\begin{figure}[H]
\label{Fig3}
\begin{center}
\includegraphics[clip,width=0.85\textwidth]{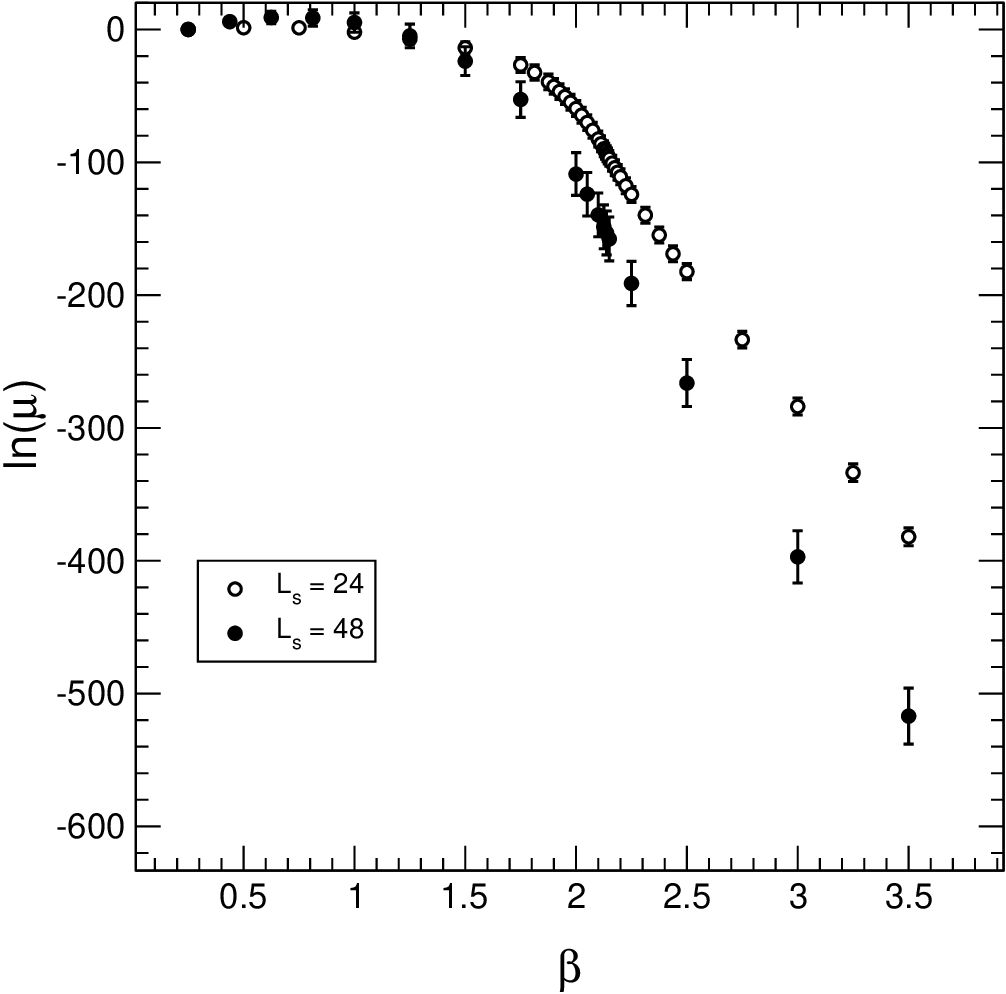}
\caption{The disorder parameter $\mu$ (Eq.~(\ref{disorder})) for
the SU(2) monopoles versus $\beta$
for lattices with $L_t=4$ and $L_s=24$ (open circles),
or $L_s=48$ (full circles).}
\end{center}
\end{figure}
\begin{figure}[H]
\label{Fig4}
\begin{center}
\includegraphics[clip,width=0.85\textwidth]{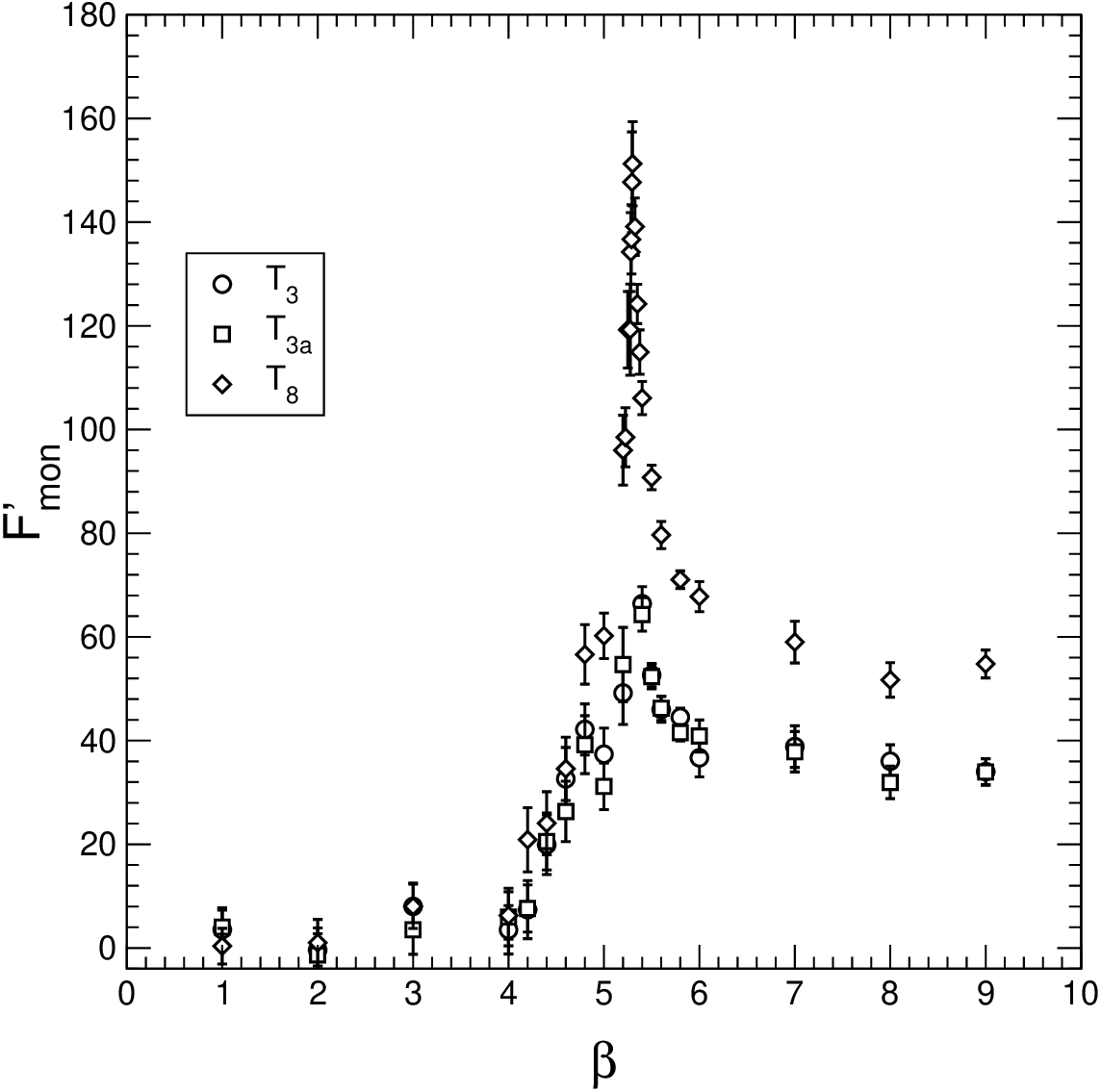}
\caption{The derivative of the SU(3) monopole free energy
in the case of $T_3$ (circles), $T_{3a}$ (squares), and $T_8$ (diamonds)
Abelian monopoles.}
\end{center}
\end{figure}
\begin{figure}[H]
\label{Fig5}
\begin{center}
\includegraphics[clip,width=0.85\textwidth]{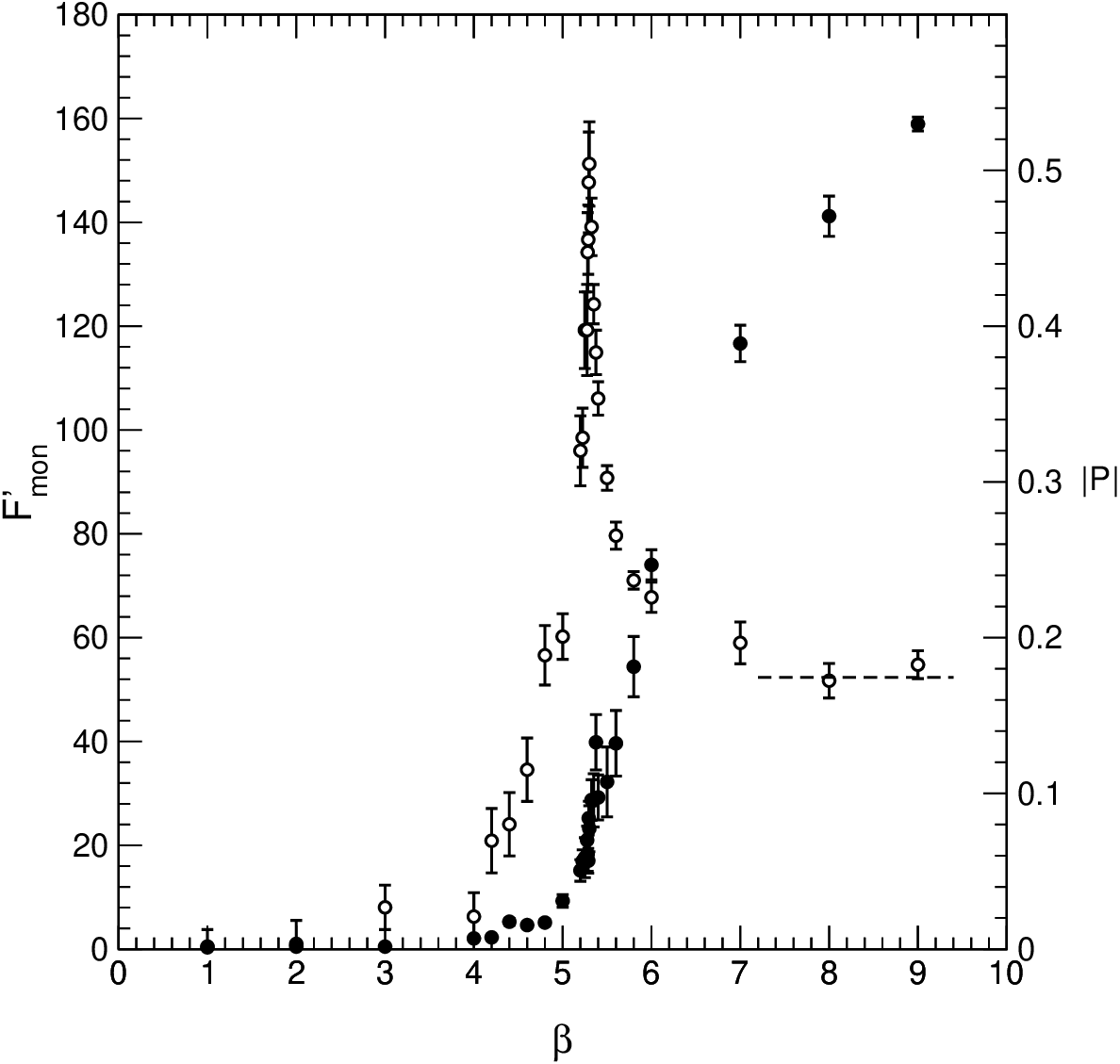}
\caption{The derivative of the SU(3) monopole free energy
for the  $T_8$ Abelian monopole (open circles) versus $\beta$ with 
the absolute value of the Polyakov loop (full circles). The dashed line
is Eq.~(\ref{classical3}).}
\end{center}
\end{figure}
\begin{figure}[H]
\label{Fig6}
\begin{center}
\includegraphics[clip,width=0.85\textwidth]{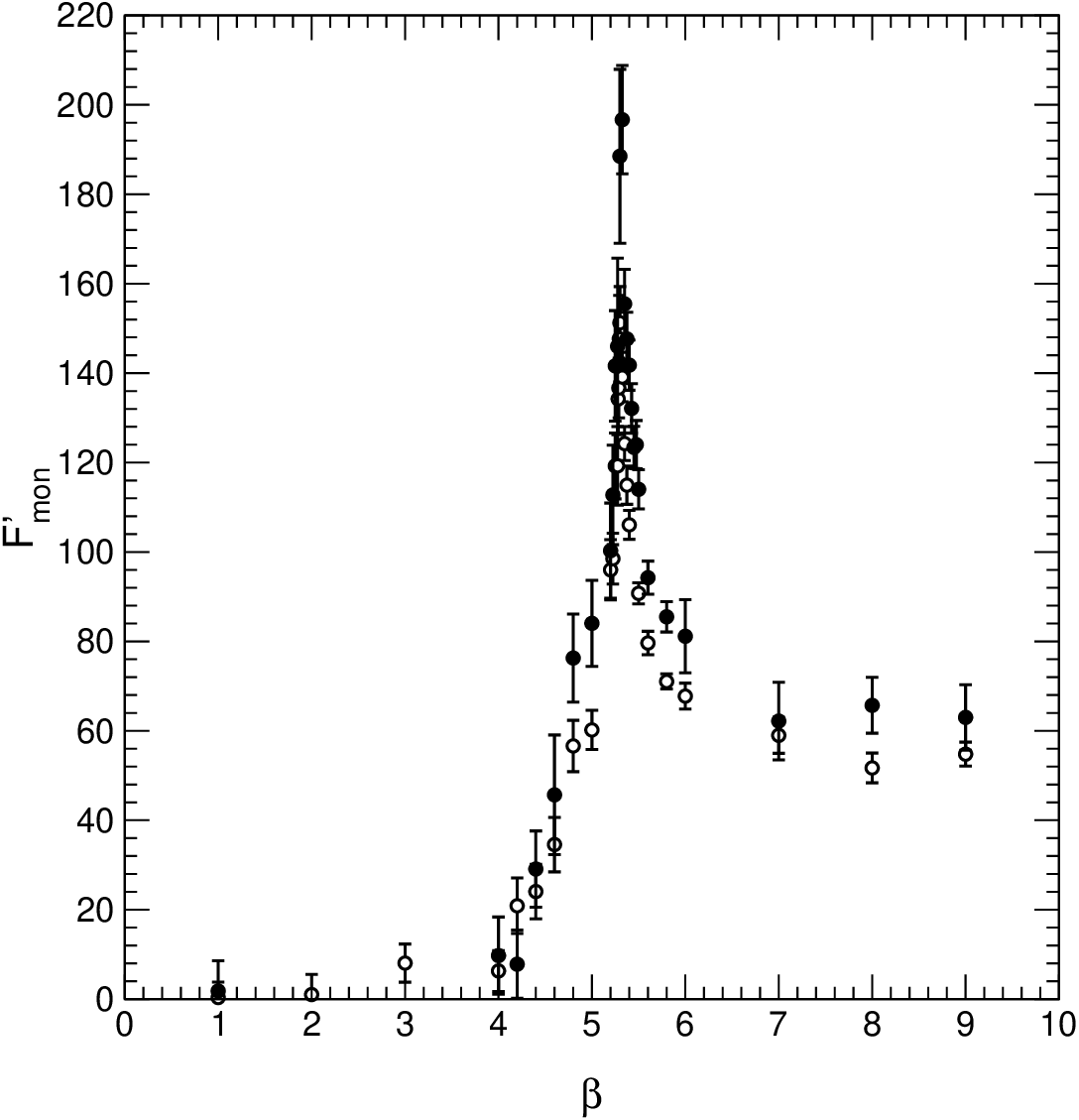}
\caption{The derivative of the $T_8$ Abelian monopole free energy versus $\beta$,
on lattices with $L_t=4$ and $L_s=24$ (open circles),
and $L_s=48$ (full circles).}
\end{center}
\end{figure}
\begin{figure}[H]
\label{Fig7}
\begin{center}
\includegraphics[clip,width=0.85\textwidth]{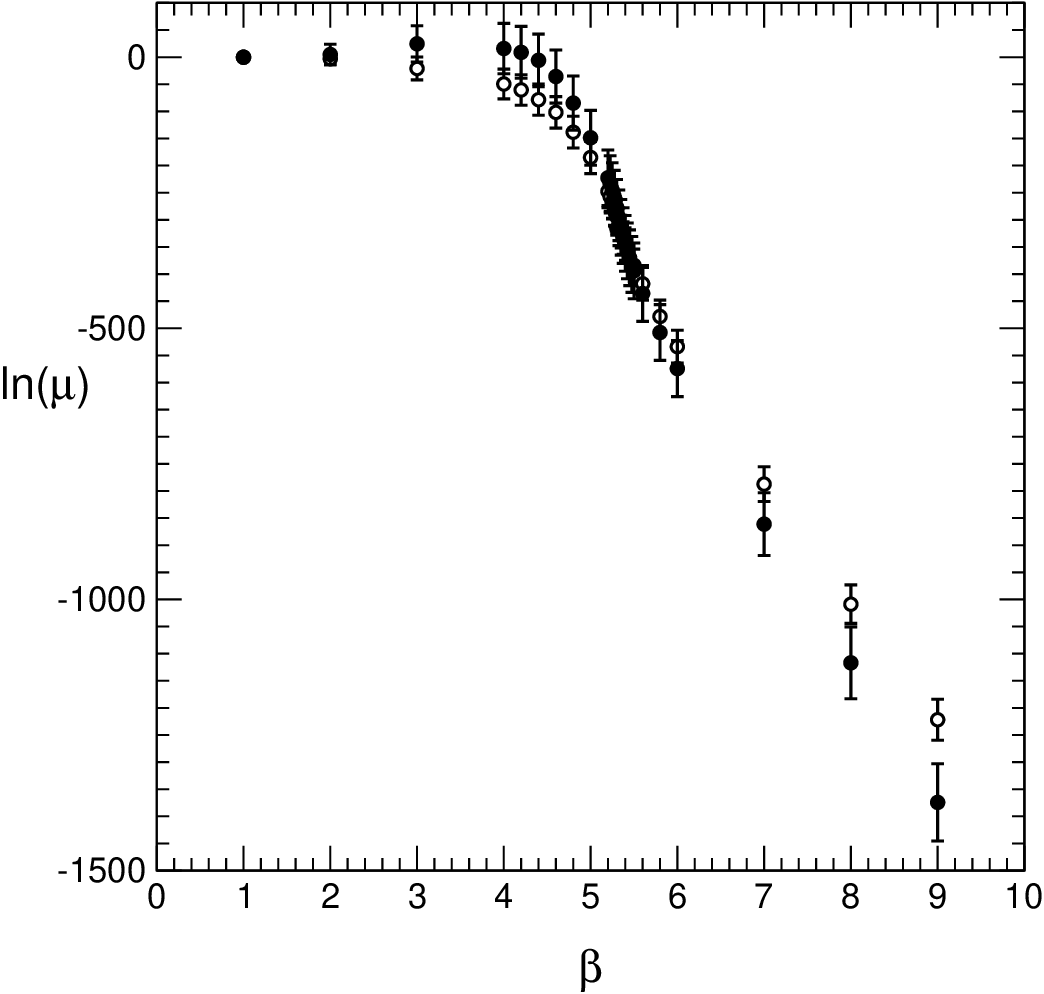}
\caption{The disorder parameter $\mu$ (Eq.~(\ref{disorder})) for
the SU(3) $T_8$ Abelian monopoles versus $\beta$
for lattices with $L_t=4$ and $L_s=24$ (open circles),
and $L_s=48$ (full circles). }
\end{center}
\end{figure}

\end{document}